\documentclass[9pt,twocolumn,twoside]{opticajnl}
\journal{opticajournal} 

\setboolean{shortarticle}{false}
\usepackage[shortcuts]{extdash} 

\title{Passive silicon nitride integrated photonics for spatial intensity and phase sensing of visible light}

\author[1,2]{Christoph Stockinger}
\author[1,2]{Jörg S. Eismann}
\author[3]{Natale Pruiti}
\author[3]{Marc Sorel}
\author[1,2,*]{Peter Banzer}

\affil[1]{Institute of Physics, University of Graz, NAWI Graz, Universitätsplatz 5, 8010, Graz, Austria}
\affil[2]{Christian Doppler Laboratory for Structured Matter Based Sensing, Universitätsplatz 5, 8010, Graz, Austria}
\affil[3]{University of Glasgow, Rankine Building, Oakfield Avenue, Glasgow G12 8LT, UK}

\affil[*]{peter.banzer@uni-graz.at}

\begin{abstract}
Phase is an intrinsic property of light, and thus a crucial parameter across numerous applications in modern optics. Various methods exist for measuring the phase of light, each presenting challenges and limitations—from the mechanical stability requirements of free-space interferometers to the computational complexity usually associated with methods based on spatial light modulators. Here, we utilize a passive photonic integrated circuit to spatially probe phase and intensity distributions of free-space light beams. Phase information is encoded into intensity through a set of passive on-chip interferometers, allowing conventional detectors to retrieve the phase profile of light through single-shot intensity measurements. Furthermore, we use silicon nitride as material platform for the waveguide architecture, facilitating broadband utilization in the visible spectral range. Our approach for fast, broadband, and spatially resolved measurement of intensity and phase enables a wide variety of potential applications, ranging from microscopy to free-space optical communication.
\end{abstract}

\setboolean{displaycopyright}{false} 

\begin{document}

\maketitle

\section{Introduction}

Light is one of the key ingredients in the evolution of modern technology.
An important contribution to this progress is made by the field of integrated photonics, which is currently undergoing rapid development \cite{Soref_2006}. 
Integrated photonics offers numerous substantial advantages, first and foremost their immense potential for miniaturization, cost\-/effectiveness at large scales, and the ability to integrate complex optical functionalities on a single chip \cite{Thomson_2016, Poon_2024}.
These developments have led to the widespread adoption of integrated photonics in various applications, including light sensing. 
In this regard, one key aspect of interest is phase-sensitive detection, which is also the central focus of this manuscript.
Phase-sensitive detection of light has many applications, including but not limited to methods of microscopy \cite{Park_2018} such as optical coherence tomography \cite{Huang_1991}, optical communication \cite{Miller_2013b, Willner_2021}, and the characterization of optical elements \cite{Book_Toeroek_2007} ranging from simple contact lenses to cutting-edge high\-/NA microscope objectives \cite{Eismann_2021a} and EUV\-/lithography optics \cite{Nomura_1999, Ma_2006}.

With a few exceptions, such as Shack\-/Hartmann wavefront sensors \cite{Platt_2001}, optical phase measurements predominantly rely on interferometry \cite{Book_Hariharan_2007}. 
Methods are generally classified as either reference\-/free or reference\-/based, with the latter requiring an external reference signal that is coherent with the light being measured \cite{Ip_2008, Rogers_2021}. 
While reference\-/based methods offer benefits, they are not always applicable, and a detailed comparison is beyond the scope of this manuscript. 
We will focus exclusively on external\-/reference\-/free methods.
In recent years, various integrated photonics based approaches have been developed for phase-resolved detection without the need for an external reference.
One approach utilizes a tree-like mesh structure of Mach\-/Zehnder interferometers, operable by power minimization\cite{Miller_2020}---easy to implement but requiring precise design specifications. 
Alternatively, the photonic mesh throughput can be analyzed numerically \cite{Buetow_2022}, accommodating imperfect optical elements at the cost of computationally expensive data evaluation. 
While both are promising, their sequential measuring routine is limited to light fields with slow temporal variations.
Other methods use a pairwise measurement scheme, which reduces complexity and enables fast readout times. 
For instance, Ref. \cite{Sun_2023} describes a narrow\-/band phase-only detection scheme.
Notably, all methods described above were realized in the near-infrared spectral range. 
While integrated photonics for visible light has existed for a long time, it faces challenges, particularly with tunable phase shifters \cite{Nejadriahi_2021}.
In addition, only recent advancements in reducing waveguide losses have made high-performance, large\-/scale photonic integrated circuits operating in the visible spectral range feasible \cite{Sacher_2019, McKay_2023}.

\begin{figure*}[tb]
    \centering
    \includegraphics[width = 1\textwidth]{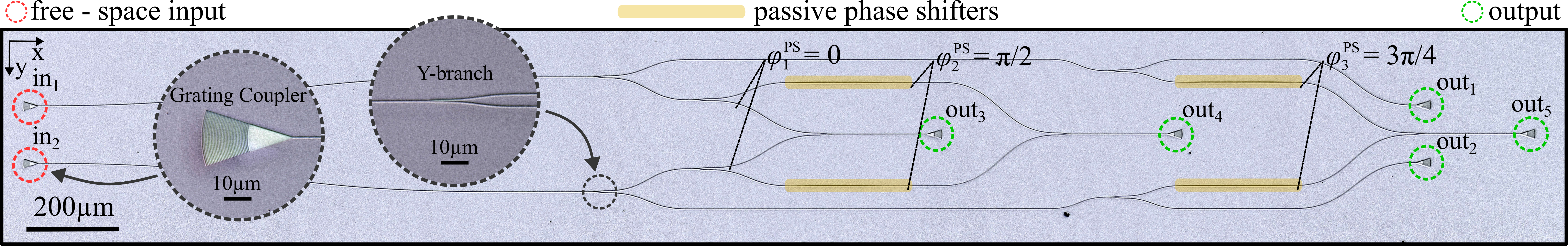}
    \caption[Microscopy image of the chip.]{Optical microscopy image of the chip. Free-space light is coupled into waveguides by means of two grating couplers. Subsequently, the signal is processed by passive on-chip interferometers. Finally, the processed light is coupled out of the chip via grating couplers again.}
    \label{fig: image-chip}
\end{figure*}

In this manuscript, we propose and experimentally verify a passive silicon nitride (SiN) photonic integrated circuit for the phase-resolved detection of visible light. 
Fundamentally, these circuits utilize a fixed set of on-chip interferometers, whose output intensity\-/only measurements allow retrieving the intensity and relative phase of the incident light.
The photonic chips route the light in a fully passive manner, eliminating the need for complex control electronics and enabling true single\-/shot measurements, with speed being only limited by the detector measuring the output intensities of the chip.
The data evaluation relies on a versatile calibration procedure that can even handle large deviations of chip elements from their design parameters, including the passive phase shifters.
As a result, even though every element of the chip is subject to chromatic dispersion, it can be accounted for through calibration, making the approach suitable for broadband applications across the visible spectrum.
Furthermore, the method employs a pairwise external-reference-free measurement scheme, offering the potential for scaling to larger detection arrays.

\section{Sensor design and measurement principle}
We start by introducing the actual integrated photonic sensor layout and the underlying design and measurement principle.
A photonic chip with two inputs is shown in Fig. \ref{fig: image-chip} (inputs marked by red circles). 
This chip was designed such that it ultimately enables retrieval of intensity and relative phase of a light field illuminating the input interface, by measuring only the intensities at the outputs (highlighted in green in Fig. \ref{fig: image-chip}) of the photonic circuit.
The input and output free-space-to-chip interface is realized via standard grating couplers \cite{4383198}.
At the input, these gratings couple the y\-/polarized component of an incident
free\-/space light field to a fundamental TE waveguide mode. 
At the output, the gratings convert the waveguide mode back into a free\-/space propagating light field, which can then be measured by external detectors.
%

On the chip, each input signal propagating along the waveguide connected to the input coupler is evenly split and directed to four waveguides using Y\-/branch splitters \cite{Zhang:13}. 
One of the four waveguides is routed directly to an output, labeled as out$_1$ and out$_2$ in Fig. \ref{fig: image-chip}, directly providing information on the intensities of the light at the corresponding input.
The remaining waveguides are connected in pairs using Y\-/branch combiner, leading to outputs labeled out$_3$, out$_4$ and out$_5$.
Two of these pairs additionally pass through a passive phase shifter before being eventually combined, which introduce a relative phase shift by altering the width of the waveguides \cite{Gonzalez-Andrade:20}. 
Together with the Y\-/branch combiners, these phase shifters and waveguide pairs form passive sampling interferometers. The outputs of these interferometers enable the calculation of the relative phase of the input light.
As previously mentioned, the intensities can easily be determined directly from the outputs out$_1$ and out$_2$, as these signals are linearly proportional to the input intensities. 
Obtaining the phase information, however, is more complex and requires a clear understanding of how the on-chip interferometers work. 
To gain the desired understanding, it is instructive to first consider a theoretical model that connects the electric fields at the inputs of the chip structure with the electric fields at its outputs.
The equations connecting the output to the input fields read:
\begin{align}
    E^\mathrm{out}_1 &= t_{11}E^\mathrm{in}_1, \label{eq: interferometer EQ1}\\
    E^\mathrm{out}_2 &= t_{22}E^\mathrm{in}_2, \label{eq: interferometer EQ2}\\
    E^\mathrm{out}_j &= t_{1j}E^\mathrm{in}_1+t_{2j}E^\mathrm{in}_2, \nonumber\\
    &= A_{1j}\mathrm{e}^{\mathrm{i}\alpha_{1j}}+A_{2j}\mathrm{e}^{\mathrm{i}\alpha_{2j}},~ &\mathrm{for~} j \mathrm{~=~ 3,4,5}. \label{eq: interferometer EQ3}
\end{align}
where $E$ represents the complex-valued electric field amplitude of the waveguide mode, while the complex proportionality coefficients $t_{ij}$ link the field at input $i$ to the field at output $j$.
The amplitude and the relative phases of the proportionality coefficients are determined using a calibration method, as detailed in Supplement 1.
 Furthermore, in \eqref{eq: interferometer EQ3}, we perform a substitution to separate the complex-valued variables into real-valued amplitude values $A_{ij} = |t_{ij}||E^\mathrm{in}_i|$ and their corresponding phase $\alpha_{ij} = \tau_{ij}+\phi_i^\mathrm{in}$, with $\tau_{ij} = \mathrm{angle}(t_{ij})$ and $\phi_i^\mathrm{in} = \mathrm{angle}(E_i^\mathrm{in})$.
 The modulus squared of \eqref{eq: interferometer EQ3} yields a well-known equation in interferometry, that clearly illustrates how the output intensity is modulated by the phase \cite{Book_Hariharan_2007}:
\begin{equation}
    I^\mathrm{out}_j = A_{1j}^2+A_{2j}^2+2A_{1j}A_{2j}\cos(\alpha_{1j}-\alpha_{2j}).
\end{equation}
We can now rearrange this equation to obtain an expression for the relative phase:
\begin{equation}
    \alpha_{1j}-\alpha_{2j} = \pm \arccos\left(\frac{I^\mathrm{out}_j-(A_{1j}^2+A_{2j}^2)}{2A_{1j}A_{2j}}  \right) + 2\pi n, \label{eq: interferometer phase}
\end{equation}
with $\alpha_{1j}-\alpha_{2j} = \tau_{1j}-\tau_{2j}+\phi_1^\mathrm{in}-\phi_2^\mathrm{in}$, and $n$ an integer number.
We drop the term $2\pi n$, since the $2\pi$ ambiguity is a common issue for interferometric phase sensors and remains unresolved in our system as well. 
Equation (\ref{eq: interferometer phase}) shows that the relative phase $\alpha_{1j}-\alpha_{2j}$ can be computed, if the output intensity of the interferometers and the amplitude factors $A_{ij}$ are known.
The values of $A_{ij}$ can be derived using equations (\ref{eq: interferometer EQ1}) and (\ref{eq: interferometer EQ2}), along with the intensities at outputs out$_1$ and out$_2$, given the proportionality $I\propto|E|^2$.
As mentioned earlier, the coefficients $t_{ij}$ are determined through a calibration process.
This calibration procedure does not determine the exact phase of the individual coefficients, it only provides information on their relative phase. 
However, since \eqref{eq: interferometer phase} exclusively depends on the relative phase of the coefficients, this information is sufficient.

The final challenge we need to address in determining the phase of the free-space light from the output signals is the fact that \eqref{eq: interferometer phase} provides two possible solutions.
To identify the correct sign of the inverse trigonometric function, additional measurements need to be performed with a known relative phase shift applied to the input signals of the interferometers.
In our case, this is done through the use of multiple interferometers with fixed phase shifters.
The correct solution can then be found as the one that is consistent across the different interferometers.
This phase reconstruction technique is often used in signal processing and is known as I/Q \-/ or In\-/phase and Quadrature technique \cite{9540747}.
Theoretically, it would suffice to have two interferometers with a non-zero difference in their preceding phase shifts. 
We, however, opt for three interferometers, as this approach adds redundancy to the system and enhances measurement accuracy.
Furthermore, the phase shifters are designed to introduce a phase delay of $\phi_{1}^\mathrm{ps}\,=\,0$, $\phi_{2}^\mathrm{ps}\,=\,\pi/2$ and $\phi_{3}^\mathrm{ps}\,=\,3\pi/4$.
These design values ensure balanced sensitivity of the device across all possible phase scenarios.
However, the phase shifts introduced on the chip can significantly deviate from their design values. Therefore, the actual phase shifts are determined through calibration and are described in terms of the relative phases of the proportionality coefficients, as explained in Supplement 1.
\section{Setup}
\begin{figure}[H]
\centering
\captionsetup{width=\linewidth}
\includegraphics[width = \linewidth]{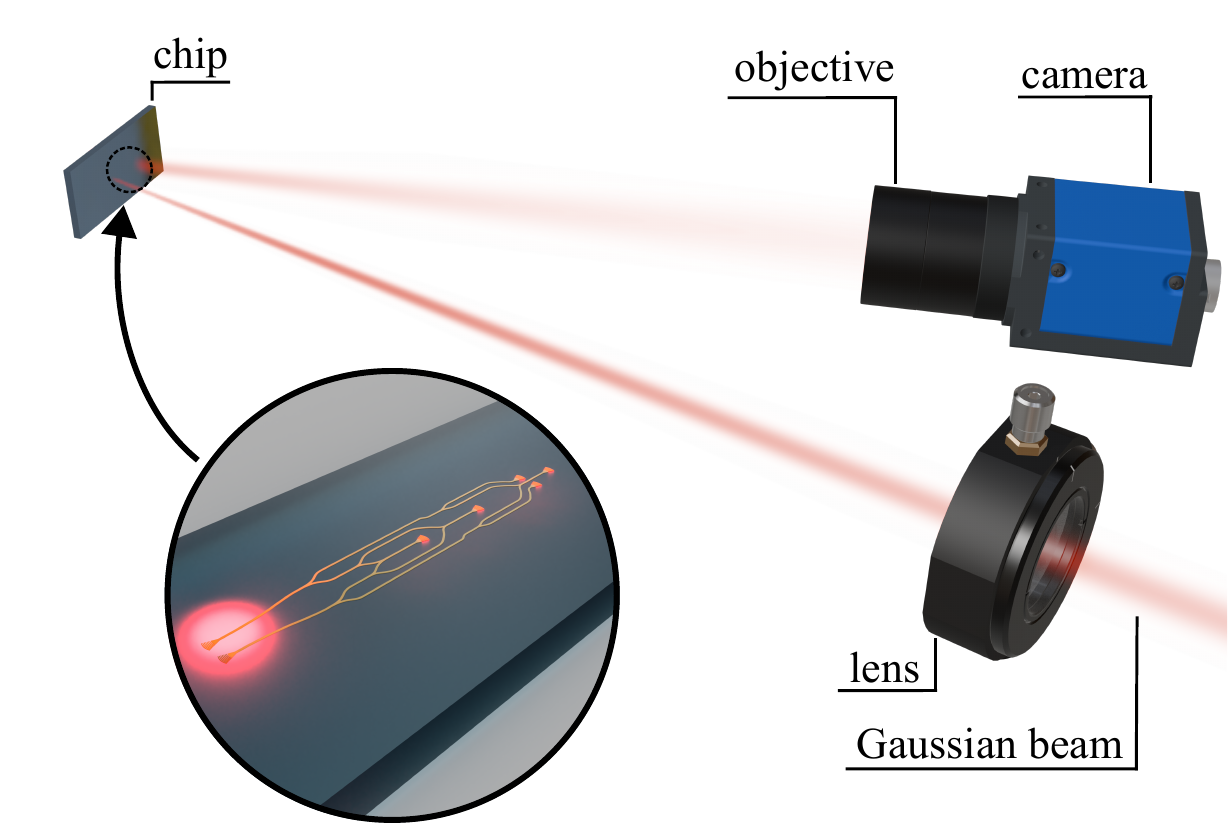}
\caption[Illustration of the experimental setup.]{Illustration of the experimental setup. A Gaussian beam is weakly focused on the input section of the chip structure. The light is coupled to waveguide modes and subsequently processed by the on chip architecture. The transmitted intensities of the outputs are monitored by means of an imaging system, which consists of a camera and an objective.}
\label{fig: setup}
\end{figure}

To investigate the proposed photonic structure, an experimental setup is required that allows for controlled illumination of the input section of the photonic circuit while simultaneously monitoring the intensity of the out-coupled light at the output section.
A schematic of the key components of the experimental setup is shown in Fig. \ref{fig: setup}.
For the experiments, we analyze light emitted by a fiber coupled laser diode (center wavelength of $\lambda$ = 658 $\mathrm{nm}$).
To maximize the efficiency of the grating couplers, which are designed to solely couple the y-polarized component of the incident light field, the polarization state is adjusted using a half\-/wave plate and a linear polarizer.
Subsequently, a lens of 400 mm focal length is placed, which weakly focuses the light onto the input region of the chip. 
The lens can be moved along the optical axis, enabling control of the beam's size and wavefront curvature at the chip position.
The chip is mounted on a 4-axis stage, allowing linear movement in three dimensions as well as adjustment of the angle of incidence of the beam with respect to the chip plane. The beam impinges on the free-space interface at an angle of 12 degrees with respect to the surface normal, which is the angle of incidence the grating couplers are designed for.
The output section of the chip is imaged onto a camera by means of an imaging system. This allows for off\-/chip monitoring of the chip's output intensity.
\section{Results and discussion}
After successfully calibrating the photonic chip using the procedure described in Supplement 1, it can be used to measure the intensity and phase of unknown free-space light fields which impinge on the input grating couplers of the system.
This solely requires a single intensity measurement at the outputs of the chip structure. 
The relative intensity at the two inputs is directly obtained using the equations (\ref{eq: interferometer EQ1}) and (\ref{eq: interferometer EQ2}). 
To determine the relative phase, \eqref{eq: interferometer phase} is used.
As previously mentioned, each interferometer provides two solutions. 
Theoretically, one could now search for a common solution for all interferometers. 
In experiments, however, it is not realistic to obtain exactly the same solution at the different interferometers. 
Instead, one searches for the solutions of the interferometers that are closest to each other, e.g. by selecting the combination of retrieved phase values that produces the smallest standard deviation. 
The average of the selected phase values of the individual interferometers is finally used as the measured relative phase.

\begin{figure}[H]
\centering
\captionsetup{width=\linewidth}
\includegraphics[width = 1\linewidth]{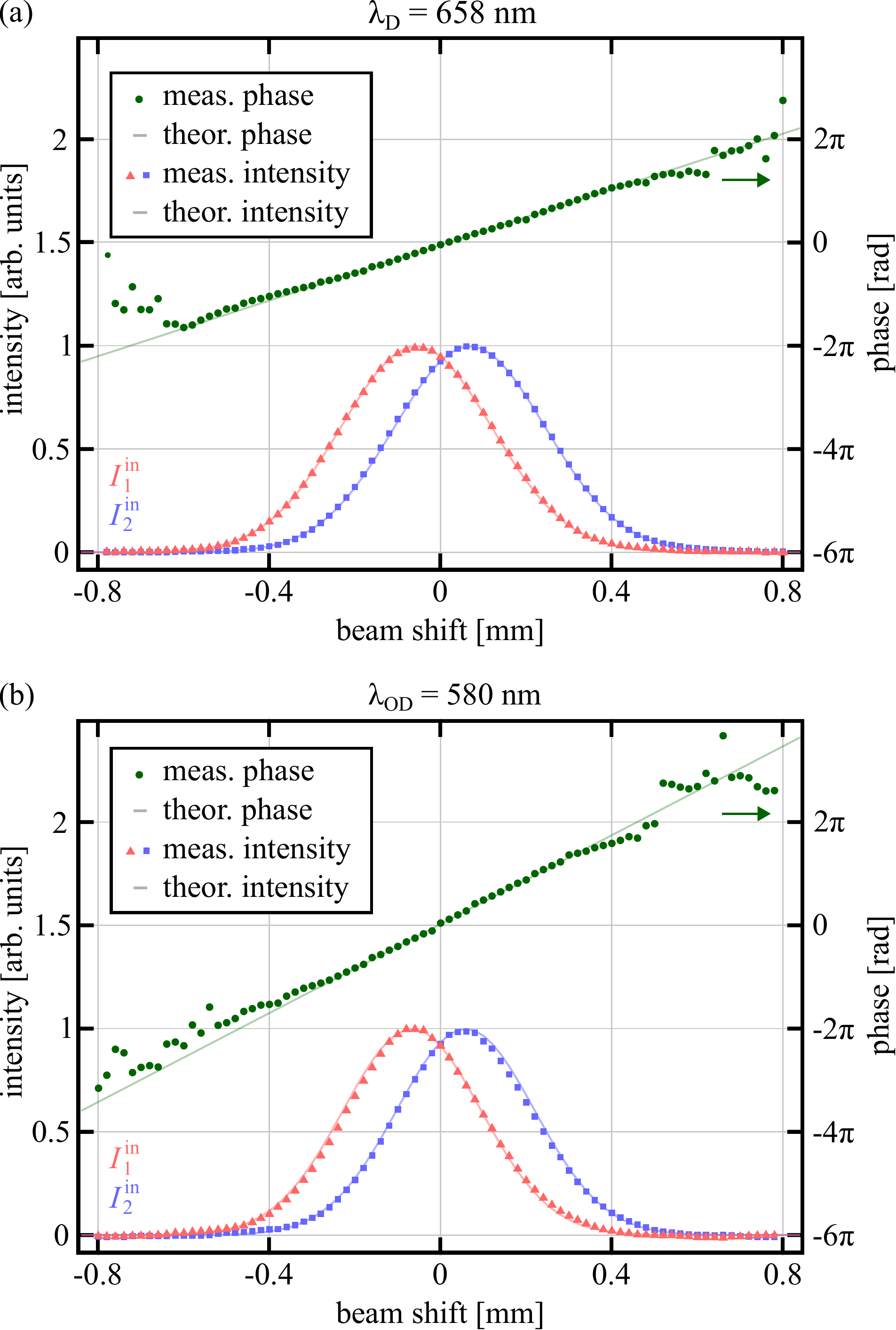}
\caption[Results]{Relative intensity and phase of a Gaussian beam as a function of the relative shift of the beam center with respect to the center of the input of the chip. (a) Measurements are performed with a beam at the design wavelength of the waveguides, $\lambda_\mathrm{D}$ = 658 nm. (b) Measurements are taken using a beam with a wavelength $\lambda_\mathrm{OD}$ = 580 nm, far from the design wavelength of the waveguides.}
\label{fig: phase and amplitude}
\end{figure}

\begin{figure}[H]
\centering
\captionsetup{width=\linewidth}
\includegraphics[width = 1\linewidth]{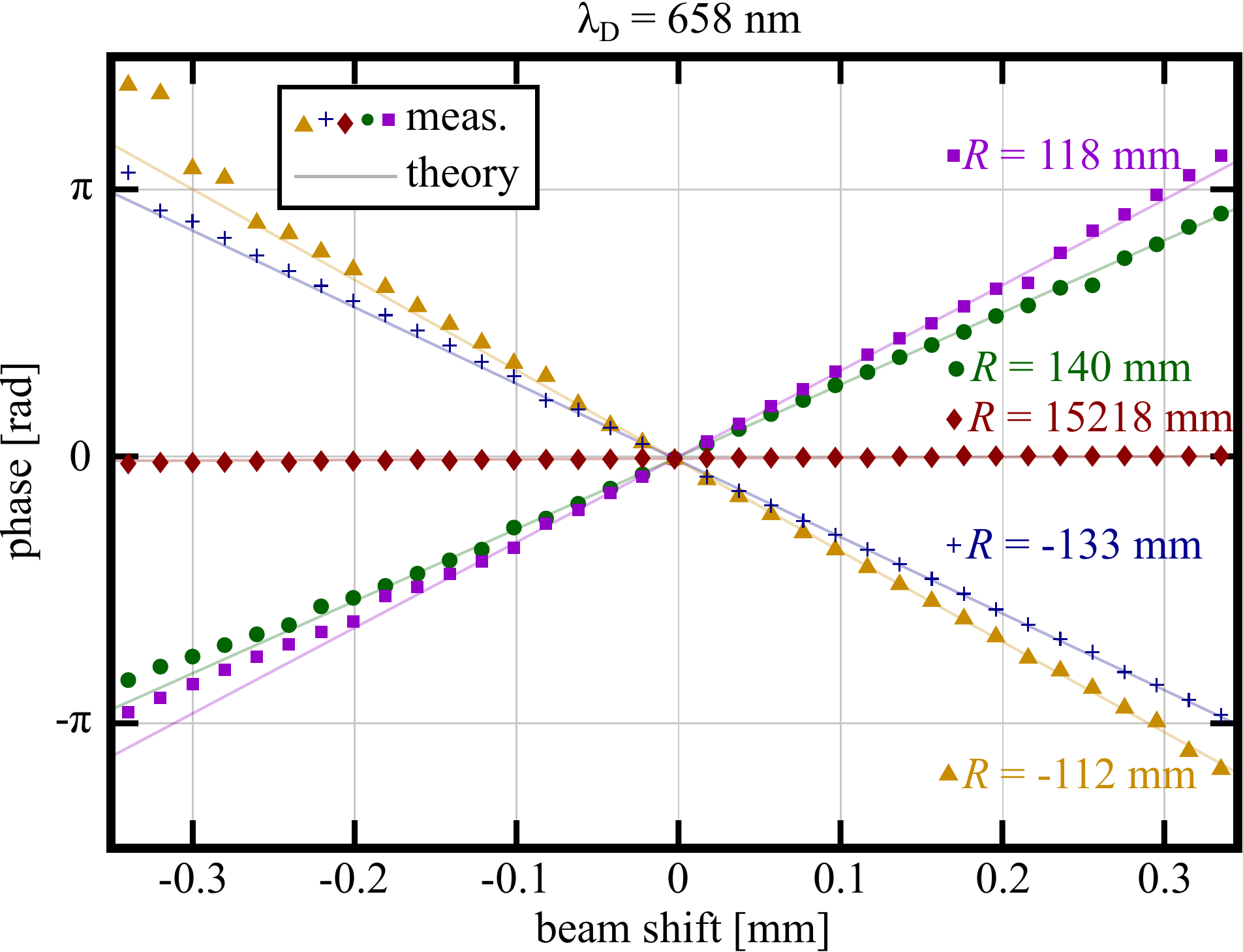}
\caption[Results]{ Relative phase of Gaussian beams of different wavefront curvature $R$. All measurements were conducted at the design wavelength of $\lambda_\mathrm{D}$ = 658 nm.}
\label{fig: phaseplot}
\end{figure}

To demonstrate the intensity and phase measurement, we scan the chip through weakly focused Gaussian beams of different parameters.
These scan measurements are very well suited to illustrate the function of the sensor, since both the relative phase and the intensity of the input signals change for different positions of the beam. 
The output intensities of the chip are recorded at each scan position individually. 
From the recorded output signals, the intensity and relative phases at the inputs are determined. 
Fig. \ref{fig: phase and amplitude} (a) shows a scan measurement of a Gaussian beam featuring a $1/e^2$ radius of $w=0.35\,\mathrm{mm}$ and a phase front curvature radius of $R = 140\,\mathrm{mm}$ at the chip surface. 
The relative phase and intensity values are plotted as a function of the relative shift of the incident beam with respect to the center of the input region of the chip. 
The theoretical values were derived from beam parameters obtained by fitting the output intensity data from the complete scan measurement.
The measured intensity reveals, as expected, the very familiar Gaussian shape, while the measured relative phase, however, is more difficult to interpret.
The relative phase of two points in space can be understood as the spatial gradient (derivative) of the phase distribution of the light beam. 
Neglecting the propagation term, and the Gouy phase, the spatial phase distribution of a paraxial Gaussian beam reads $\phi = k\frac{r^2}{2R}$, where $k$ is the wave number and $r$ the radial distance to the beam center. 
Note, that its spatial gradient is a linear function of the radial position $r$, with a slope inversely proportional to the radius of the phase front curvature $R$, explaining the linear trend seen in the measurement.
The linear behavior of the relative phase described above is confirmed by the measurements shown in Fig.\ref{fig: phaseplot}, where the relative phase of Gaussian beams with phase fronts of different curvature is plotted as a function of the relative shift of the beam with respect to the center of the coupling region of the chip.
Note that for all the results presented, each amplitude and phase value is derived from individual measurements.
The data is plotted with a common x-axis to simplify interpretation and illustrate a certain systematic pattern. 
However, the results presented can be considered individual measurements that demonstrate the functionality of the circuit across a variety of different scenarios.
The calibration process not only compensates for manufacturing inaccuracies, it also accounts for the chromatic behavior of the on\-/chip components.
As a result, the chips can be effectively used at wavelengths far from their design value, once calibrated for the desired wavelength.
%
It should be noted that the chip can be calibrated for operation at any wavelength, provided the waveguides and grating couplers are sufficiently efficient. Additionally, the waveguides must remain single-mode, as this is essential for the proper functioning of the Y\-/branches.
If higher-order modes are excited, the theoretical model discussed for the chip is no longer applicable, causing the measurement principle to fail.

To showcase the broadband capabilities of the presented chip design, scan measurements of non\-/collimated Gaussian beams were performed at a wavelength of $\lambda_\mathrm{OD} = 580$ nm, significantly different from the design wavelength of $\lambda_\mathrm{D} = 658$ nm.
Fig. \ref{fig: phase and amplitude} (b) shows a scan measurement of a Gaussian beam featuring a $1/\mathrm{e}^2$ radius of $w = 0.32$ mm and a phase front curvature radius of $R = 100$ mm.
As with the previous measurements at the design wavelength, there is excellent agreement between the experimental results and theory.

\begin{figure*}[tb]
\centering
\includegraphics[width = 1\textwidth]{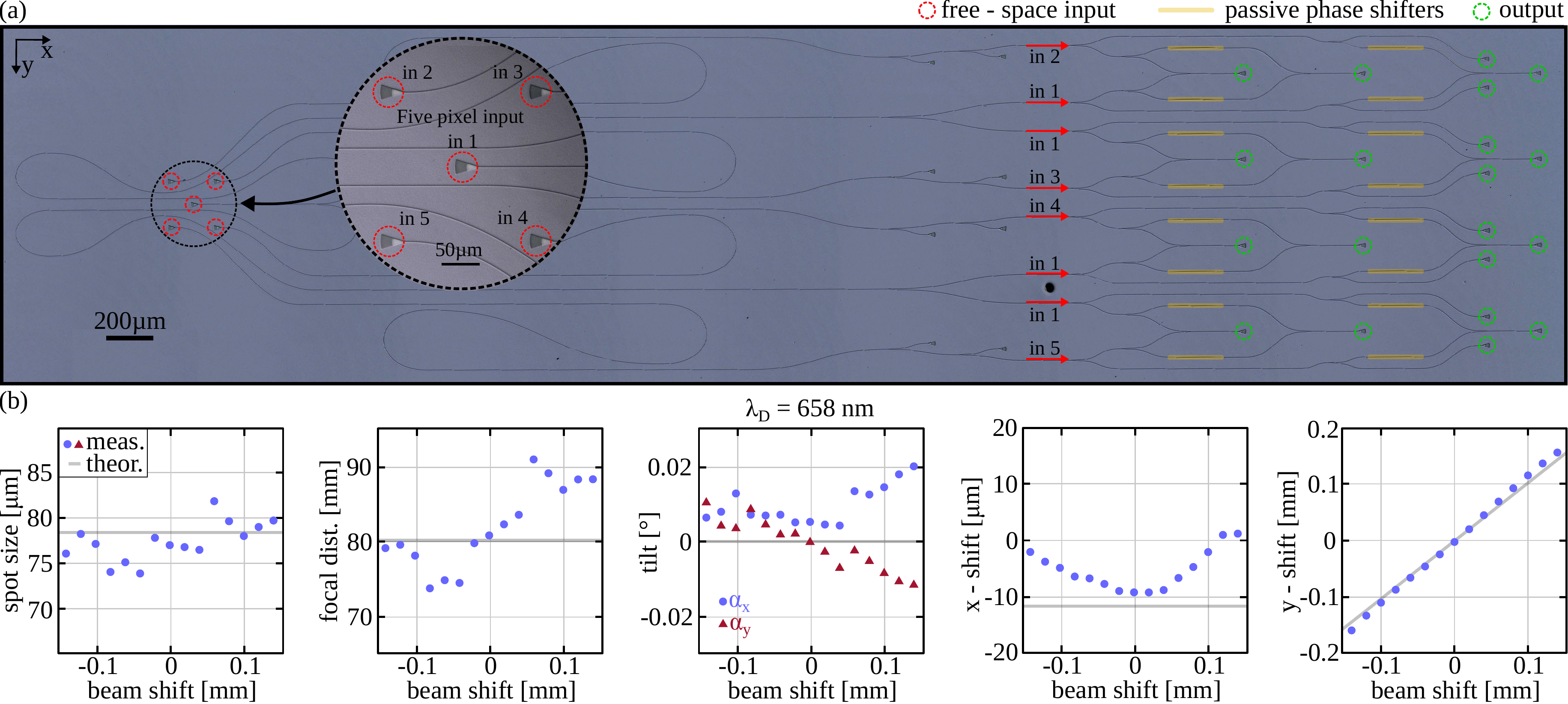}
\caption[Microscopy image of the chip featuring 5 input pixels.]{(a) Optical microscope image of the chip featuring a 5\-/pixel input interface for demonstration of scalability. The input interface consists of five grating couplers functioning as input pixels, arranged in a square configuration with four corner pixels and one central pixel. The on-chip architecture is designed such that each corner pixel is connected to the central pixel via a phase and intensity measurement unit. This design facilitates the complete characterization of a Gaussian beam and its parameters through a single-shot intensity measurement at the outputs. (b) Retrieved parameters of a Gaussian beam as a function of the relative displacement of the center of the beam with respect to the center of the input section.}
\label{fig: image-5pix-chip}
\end{figure*}
\section{First steps towards larger structures}
After having demonstrated experimentally the capabilities of the passive photonic circuit with respect to phase and intensity measurements, we now discuss a chip architecture featuring more input pixels and show corresponding measurement results.
To showcase the scalability of our system, we designed and fabricated a photonic chip with a 5\-/pixel input interface.
A microscope image of the chip is shown in Fig. \ref{fig: image-5pix-chip} (a).
Again, focussing grating couplers are used as free-space interfaces.
The five couplers are arranged in a square, with four pixels at the corners and a fifth in the center.
Each of the corner pixels is connected to the central pixel using a phase and intensity measuring unit, similar to the on-chip architecture discussed earlier.
The pairwise phase and intensity measurements in the five-pixel chip follow the same principles as the previously discussed two-pixel architectures.
This similarity allows the established calibration method to be applied again without any conceptual modifications.
Moreover, this specific pixel arrangement not only facilitates the measurement of the relative phase and intensity between the corner pixels and the central pixel, but also enables the reconstruction of the parameters of a paraxial Gaussian beam through a single-shot measurement of the output intensities.
A detailed description of the reconstruction of the beam parameters from measured intensity and phase data is provided in Supplement 1.

Fig. \ref{fig: image-5pix-chip} (b) shows retrieved parameters of a Gaussian beam featuring a $1/\mathrm{e}^2$ radius of $w$ = 0.23 mm and a phase front curvature radius of $R$ = 91 mm at the chip surface. We analyze the spot size, focal distance, tilt angles in the x- and y-directions, as well as the beam shift in the x- and y-directions.
The parameters are presented for various y-positions of the beam relative to the center of the chip's input.
It can be seen that the reconstruction of the beam parameters performs well for slight misalignment between the beam and the chip.
Although a decrease in accuracy is evident with increasing misalignment, it is important to note that in the data shown, some pixels receive less than $1/e^2$ of the maximum intensity at a misalignment of 150 µm.
Reduced input intensity leads to a decreased signal-to-noise ratio when measuring the output signals, resulting in less accurate phase and parameter reconstruction.
Nevertheless, the data demonstrates that the parameters of a Gaussian beam can be accurately determined using a single-shot intensity measurement. 
Furthermore, the data indicates a significant level of insensitivity of the method with respect to misalignment.
\section{Conclusion}
A photonic integrated circuit capable of spatially resolving phase and intensity of visible free-space light has been proposed and experimentally demonstrated.
The chip utilizes a fixed set of passive on-chip interferometers, whose output intensity measurements enable the retrieval of the intensity and relative phase information of the incident light field.
The capabilities of the circuit have been demonstrated through scan measurements of uncollimated Gaussian beams of varying parameters.
Additionally, the potential for broadband application of the structure has been showcased through measurements conducted at different wavelengths, specifically $\lambda_\mathrm{D} = 658$ nm and $\lambda_\mathrm{OD} = 580$ nm.
Finally, first steps toward larger structures were discussed. A chip featuring five input pixels was presented, and its extended functionality was demonstrated by reconstructing all parameters of a paraxial Gaussian beam from single-shot measurements of its output intensities.

Notably, recent advancements in integrated photonics could be incorporated into the presented chip design to enhance functionality and integration. 
Potential modifications include a more expansive and generic input interface, polarization splitting grating couplers \cite{mi11070666,su2018fully} for resolving also light's polarization, and on-chip photodiodes \cite{DeVita:22}.

The presented approach and the actual integrated photonic system constitute a powerful, versatile, and small-footprint addition to the existing toolboxes of light field metrology.

\begin{backmatter}

\bmsection{Acknowledgments}
The financial support by the Austrian Federal Ministry of Labour and Economy, the National Foundation for Research, Technology and Development and the Christian Doppler Research Association is gratefully acknowledged.

\bmsection{Disclosures}
The authors declare no conflicts of interest.






\bmsection{Data availability} Data underlying the results presented in this paper are not publicly available at this time, but may be obtained from the authors upon reasonable request.

\bigskip

\bmsection{Supplemental document}
See Supplement 1 for supporting content. 

\end{backmatter}

\bibliography{literature}

\end{document}


\maketitle

\section{Calibration}
\label{sec:calibration}

Before conducting measurements, it is essential to determine all the unknown parameters of the chip elements through precise calibration procedure. 
The calibration is based on illumination of the chip with known light fields, the output of the chip then provides information about the behaviour of the chip components.   
From this information, we determine proportionality coefficients of the form $t_{ij} = |t_{ij}| e^{\mathrm{i} \tau_{ij}}$ that account for both losses and phase shifts introduced as light propagates from input $i$ to output $j$. 
Using these proportionality coefficients, we can derive equations that link the input fields to the output fields of the chip:
\begin{align}
    E^\mathrm{out}_1 &= t_{11}E^\mathrm{in}_1, \label{eq: loss-calib-1}\\ 
    E^\mathrm{out}_2 &= t_{22}E^\mathrm{in}_2, \label{eq: loss-calib-2}\\
    E^\mathrm{out}_j &= t_{1j}E^\mathrm{in}_1 + t_{2j} E^\mathrm{in}_2~,~ \mathrm{for}~ j~\mathrm{ = ~3,~ 4,~ 5}.
    \label{eq: loss-calib-3}
\end{align}
Whereby the amplitudes of the complex coefficients $|t_{ij}|$ describe all the losses that are introduced by imperfection of components. The phase of the coefficients, on the other hand, describes the phase shifts introduced in the waveguides. These include parasitic phase shifts resulting from manufacturing inaccuracies, as well as intentional phase shifts introduced by the phase shifters.
The calibration process for determining $t_{ij}$ consists of two steps for the selected chip design.

In the first step, we determine the amplitude of $t_{ij}$.
This is done by individually illuminating the input couplers with light of uniform intensity distribution.
If only one input is exposed with light, the output intensities provide information on the relative power throughput from the exposed input to the outputs. 
The normalized output intensities directly give information about the amplitude of the proportionality coefficients.

In the second step, we aim to determine the phase information of the proportionality coefficients.
%
%
%
This is done by illuminating the input section of the chip by means of a light field of known amplitude and phase distribution.
All inputs are now exposed simultaneously, allowing light to enter both input arms of the on-chip interferometers.
The relative phase of the waveguide signals determines the output intensity of the interferometers.
The phase distribution of the input light is known, allowing the output signals from the interferometers to reveal the additional relative phase shifts introduced by the chip.
However, measuring just a single set of amplitude and phase scenarios at the  input pixels is not enough to determine the phase shifts accurately, as there is an ambiguity in the sign when extracting the phase from the interference signal.
Theoretically, this issue can be resolved by conducting at least two measurements with different scenarios at the inputs.
In practice, many more than two measurement points are collected, primarily to enhance the accuracy of the calibration and to account for the additional unknown parameters that need to be determined in this process.
There are a variety of options for generating different scenarios at the inputs. 
The option of our choice is to scan a non-collimated Gaussian beam, as it features a curvature in its phase front, and therefore provides different input phases for different scan positions.
This allows us to describe the entire system using the equations \eqref{eq: loss-calib-1} to \eqref{eq: loss-calib-3} with $E_1^\mathrm{in}$ and $E_2^\mathrm{in}$ as the electric field of the Gaussian beam at the positions of the input grating couplers.
The theory model can then be fitted to measured data, using the relative phase of the proportionality coefficients $\Delta \tau_j = \tau_{1j}-\tau_{2j}$, among other variables, as free parameters. 
The relative phases $\Delta \tau_j$ can therefore be simply extracted from the fitted model.
As described above, only the phase differences of the coefficients that describe the behavior of the on-chip interferometers is determined.
However, this is sufficient because only these relative phases are needed for subsequent measurements with the chip.

To ensure reproducibility and determine systematic errors, this calibration procedure was repeated a total of eight times with beams of four different parameters. 
This analysis revealed that the on-chip phase shifts exhibit completely random values, deviating significantly from the design parameters. 
However, the calibration procedure proved to be very robust, with the standard deviation of the determined phase values across different calibration scenarios being only $\sigma_{\Delta\tau, \mathrm{min}} = 0.010 \, \mathrm{rad}$ to $\sigma_{\Delta\tau, \mathrm{max}} = 0.018 \, \mathrm{rad}$.

\section{Characterizing a paraxial Gaussian beam from
phase and intensity information of five pixels}
\label{sec: Sup Characterizing Gaussian beam}
Before we proceed with the calculation of the beam parameters from intensity and phase data obtained with the five pixel architecture, it is important to first discuss the simplifications applied in the calculation.

First, it should be noted that, due to the design of the grating coupler, the chip is operated at an angle to the optical axis.
This means that the chip is calibrated to a specific angle of incidence, so that if a plane wave strikes the chip at this angle, the phase measurement would yield zero relative phase between all inputs. 
However, it also means that different input pixels are situated in different planes along the direction of beam propagation.
As the wavefront curvature and the Gouy phase of non-collimated Gaussian beams change during propagation, this results in different measured phase values depending on the position of the pixel.
Nevertheless, since we only consider very weakly focused beams that are examined relatively far from their focus, and given the small angle of incidence and distance between the grating couplers, this phase contribution is assumed to be negligible and is therefore not treated in our model.
Additionally, we only consider beams that hit the chip's input at an angle that deviates only slightly from the calibration angle.
Consequently, we can assume that the change in the distance of the pixels to the optical axis due to tilt is negligible.
With all these assumptions, we can build a mathematical model as if the chip surface is placed perpendicular to the optical axis, simplifying calculations significantly.
Note that all these assumptions introduce systematic errors in the reconstruction of beam parameters. 
Nevertheless, the model performs well enough to demonstrate the reconstruction of paraxial Gaussian beams.

The mathematical description begins with a very simple representation of the electric field of a paraxial Gaussian beam at the position of the chip at the optical axis. Here we neglect polarization and the time harmonic oscillation, leading to the following expression \cite{svelto2010principles}:

\begin{equation}
    E(r,z) = \frac{w_0}{w}E_0e^{-\frac{r^2}{w^2}}\mathrm{e}^{-\mathrm{i}\left(kz+k\frac{r^2}{2R}-\Psi\right)}, \label{eq: Gaussian Beam}
\end{equation}
with $E_0$ as the amplitude at the beam origin, $r$ the radial distance to the center of the beam, $w$ the beam size at the position of the chip, $R$ the radius of the wavefront curvature at the position of the chip, $\Psi$ the Gouy phase, $w_0$ the spot size at the focal point and $k$ the wavenumber.

\begin{figure}[H]
\centering
\includegraphics[width = 0.3\linewidth]{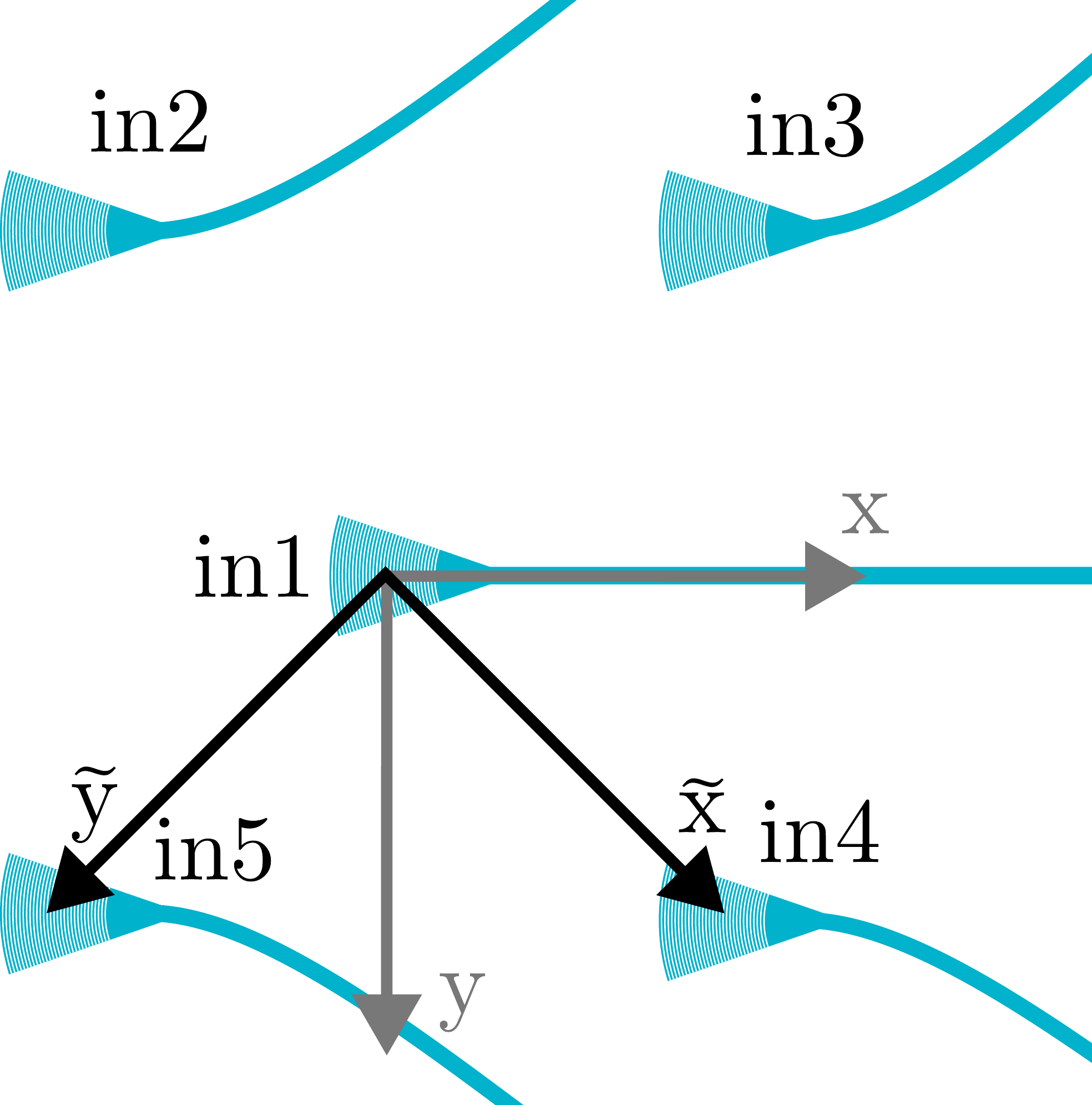}
\caption[]{Illustration of the coordinate transformation.}
\label{fig: coordinate transformation}
\end{figure}

We can now separate this equation into two parts to analyze the amplitude and phase independently.
Starting with the amplitude parts, we can express measured input intensity values at input $i$ using the proportionality $\sqrt{I} \propto E$:
\begin{equation}
    \sqrt{I^\mathrm{in}_i} \propto \frac{w_0}{w}E_0\mathrm{e}^{-\left(\frac{r^\mathrm{in}_i}{w}\right)^2}.
    \label{eq: Gaussian Beam Amplitude}
\end{equation}
The measured intensities at the positions of the input pixels provide information about the relative position of the beam in the plane of the chip and its size $w$.
To retrieve these parameters, we first represent the radial distance $r$ at the position of the input pixel $i$ in Cartesian coordinates:
\begin{equation}
    (r_i^\mathrm{in})^2 = (x_i^\mathrm{in} - x_0)^2 + (y_i^\mathrm{in} - y_0)^2,
\end{equation}
with $x_0$ and $y_0$ as the displacements of the beam with respect to the center of the input section of the chip.
Next, we rotate the coordinate system by 45°, as illustrated in Fig. \ref{fig: coordinate transformation}, which will later demonstrate a significant simplification in the calculations.
The new coordinates read as follows:
\begin{align}
    \Tilde{x}_i^\mathrm{in} &= \frac{1}{\sqrt{2}}(x_i^\mathrm{in}+y_i^\mathrm{in}),
    \label{eq: coordinate transform 1}\\
    \Tilde{y}_i^\mathrm{in} &= \frac{1}{\sqrt{2}}(-x_i^\mathrm{in}+y_i^\mathrm{in}).
    \label{eq: coordinate transform 2}
\end{align}
By substituting the new coordinate system into \eqref{eq: Gaussian Beam Amplitude}, we can derive the following two equations:
\begin{align}
    \sqrt{\frac{I_1^\mathrm{in}}{I_3^\mathrm{in}}} &= \mathrm{e}^{-\frac{1}{w^2}\left[\Tilde{y}_0^2-(\Tilde{y}_3^\mathrm{in}-\Tilde{y}_0)^2\right]},
    \label{eq:y_0-1}\\
    \sqrt{\frac{I_1^\mathrm{in}}{I_5^\mathrm{in}}} &= \mathrm{e}^{-\frac{1}{w^2}\left[\Tilde{y}_0^2-(\Tilde{y}_5^\mathrm{in}-\Tilde{y}_0)^2\right]}.
    \label{eq:y_0-2}
\end{align}
Isolating $w$ in equation \eqref{eq:y_0-1} and inserting it into equation \eqref{eq:y_0-2}, as well as using that, $\Tilde{y}_3^\mathrm{in} = - \Tilde{y}_5^\mathrm{in}$, we can find an expression for the displacement in $\Tilde{y}$-direction:
\begin{equation}
    \Tilde{y}_0 = \frac{\Tilde{y}_3^\mathrm{in}\left[\ln\left(\frac{I_1^\mathrm{in}}{I_5^\mathrm{in}}\right)-\ln\left(\frac{I_1^\mathrm{in}}{I_3^\mathrm{in}}\right)\right]}{2\left[\ln\left(\frac{I_1^\mathrm{in}}{I_5^\mathrm{in}}\right)+\ln\left(\frac{I_1^\mathrm{in}}{I_3^\mathrm{in}}\right)\right]}.
\end{equation}
Repeating this procedure with the intensities at inputs 1, 2, and 4 reveals the displacement in the $\Tilde{x}$-direction:
\begin{equation}
    \Tilde{x}_0 = \frac{\Tilde{x}_3^\mathrm{in}\left[\ln\left(\frac{I_1^\mathrm{in}}{I_4^\mathrm{in}}\right)-\ln\left(\frac{I_1^\mathrm{in}}{I_2^\mathrm{in}}\right)\right]}{2\left[\ln\left(\frac{I_1^\mathrm{in}}{I_4^\mathrm{in}}\right)+\ln\left(\frac{I_1^\mathrm{in}}{I_2^\mathrm{in}}\right)\right]}.
\end{equation}
Knowing $\Tilde{x}_0$ and $\Tilde{y}_0$ we can now calculate the radial distance of each pixel to the beam center $r_i^\mathrm{in}$. 
Furthermore, by rearranging \eqref{eq:y_0-1} we can find an expression for the beam size $w$, measured in the $\Tilde{y}$-direction:
\begin{equation}
    w_\mathrm{\Tilde{y}} = \sqrt{\frac{-2\left[(r^\mathrm{in}_1)^2-(r^\mathrm{in}_3)^2\right]}{\ln\left( \frac{I_1^\mathrm{in}}{I_3^\mathrm{in}}\right)}}.
\end{equation}
The same procedure can be applied to the equations for the intensities of the inputs 1 and 4 to obtain an expression for the beam size, measured in the  $\Tilde{x}$-direction:
\begin{equation}
    w_\mathrm{\Tilde{x}} = \sqrt{\frac{-2\left[(r^\mathrm{in}_1)^2-(r^\mathrm{in}_4)^2\right]}{\ln\left( \frac{I_1^\mathrm{in}}{I_4^\mathrm{in}}\right)}}.
\end{equation}
Using the measured intensity values, we can determine both the in-plane displacement and the size of the beam at the position of the chip.

Next, we aim to determine the curvature of the wavefront and the tilt of the beam, which requires a closer examination of the measured relative phase values.
The relative phase between two inputs i and j can be expressed as follows:
\begin{equation}
    \Delta\phi_{ij}^\mathrm{in} = k\frac{(r^\mathrm{in}_i)^2-(r^\mathrm{in}_j)^2}{2R}\pm \phi_\mathrm{tilt},
    \label{eq: relative phase}
\end{equation}
with $\Delta\phi_{ij}^\mathrm{in} =\phi_i^\mathrm{in}-\phi_j^\mathrm{in}$.
We assume that both inputs experience the same Gouy phase. Furthermore, we reformulate the propagation terms of the relative phase to a phase term $\phi_{tilt}$ that appears when a tilt between the chip surface normal and the optical axis is present.
Initially, we incorporate both signs for the tilt term in \eqref{eq: relative phase}, as the sign of the tilt term depends on the relative positions of the two couplers being analyzed.
By adding \eqref{eq: relative phase} for the relative phases $\Delta\phi_{13}^\mathrm{in}$ and $\Delta\phi_{15}^\mathrm{in}$ and isolating 
$R$, we can derive a representation for the radius of the wavefront curvature in the $\Tilde{y}$-direction:
\begin{equation}
    R_{\mathrm{\Tilde{y}}} = \frac{k \left( 2(r^\mathrm{in}_1)^2 - \left[(r^\mathrm{in}_3)^2 + (r^\mathrm{in}_5)^2\right] \right)}{2\left(\Delta\phi_{13}^\mathrm{in}+\Delta\phi_{15}^\mathrm{in}\right)}.
\end{equation}
Similarly, we can perform the same process for the relative phases $\Delta\phi_{12}^\mathrm{in}$ and $\Delta\phi_{14}^\mathrm{in}$ to derive an expression for the radius of the wavefront curvature in the \(\tilde{x}\)-direction:
\begin{equation}
    R_{\mathrm{\Tilde{x}}} = \frac{k \left( 2(r^\mathrm{in}_1)^2 - \left[(r^\mathrm{in}_2)^2 + (r^\mathrm{in}_4)^2\right] \right)}{2\left(\Delta\phi_{12}^\mathrm{in}+\Delta\phi_{14}^\mathrm{in}\right)}.
\end{equation}
Using an analogous procedure, we can ultimately derive the representation for the tilt terms associated with tilting around the \(\tilde{x}\)- and \(\tilde{y}\)-axes:
\begin{equation}
    \phi_\mathrm{tilt}^\mathrm{\Tilde{x}} = \frac{1}{2} \left( \Delta\phi_{15}^\mathrm{in}-\Delta\phi_{13}^\mathrm{in} \right) + \frac{\left(-(r^\mathrm{in}_3)^2 + (r^\mathrm{in}_5)^2\right)\left( \Delta\phi_{13}^\mathrm{in}+\Delta\phi_{15}^\mathrm{in}\right)}{2\left(2(r^\mathrm{in}_1)^2 - (r^\mathrm{in}_3)^2 - (r^\mathrm{in}_5)^2\right)},
\end{equation}
\begin{equation}
    \phi_\mathrm{tilt}^\mathrm{\Tilde{y}} = \frac{1}{2} \left( \Delta\phi_{14}^\mathrm{in}-\Delta\phi_{12}^\mathrm{in} \right) + \frac{\left(-(r^\mathrm{in}_2)^2 + (r^\mathrm{in}_4)^2\right)\left( \Delta\phi_{12}^\mathrm{in}+\Delta\phi_{14}^\mathrm{in}\right)}{2\left(2(r^\mathrm{in}_1)^2 - (r^\mathrm{in}_2)^2 - (r^\mathrm{in}_4)^2\right)}.
\end{equation}
The tilt angles between the structure and the incident beam, associated with the measured phase values, can subsequently be determined using straightforward geometric considerations.

Above, we discussed the parameters of a paraxial Gaussian beam at the chip position. 
However, for a comprehensive characterization of the beam, it is also interesting to know its focal spot size $w_0$ and the distance to its focal plane $z$.
These two parameters can be calculated using the previously determined beam size and wavefront curvature.
To accomplish this, we first describe the beam size and wavefront curvature with the following expressions \cite{svelto2010principles}:
\begin{align}
    R(z) &= z\left(1-\left(\frac{z_\mathrm{R}}{z}\right)^2\right),
    \label{eq: wavefront curvature}\\
    w(z) &= w_0\sqrt{1+\left(\frac{z}{z_\mathrm{R}}\right)},
    \label{eq: beam size}
\end{align}
where $z_\mathrm{R}$ is the Rayleigh range, which is defined as follows:
\begin{equation}
    z_\mathrm{R} = \frac{\pi w_0^2 n}{\lambda},
\end{equation}
with the refractive index of the medium $n$ and the wavelength $\lambda$.
We can now substitute the average of the previously determined wavefront curvature and beam size, with $ R_\mathrm{c} = \frac{R_{\mathrm{\Tilde{x}}}+R_{\mathrm{\Tilde{y}}}}{2}$ and $ w_\mathrm{c} = \frac{w_{\mathrm{\Tilde{x}}}+w_{\mathrm{\Tilde{y}}}}{2}$.
Transforming and substituting \eqref{eq: wavefront curvature} and \eqref{eq: beam size}, we can then derive expressions for $z$ and $w_0$.
\begin{align}
    z &= \frac{R_c (w_c^2 \pi n)^2}{\lambda^2 R_c^2 + (w_c^2 \pi n)^2},\\
    w_0 &= \frac{R_cw_c}{\sqrt{R_c^2+\left(\frac{w_c^2\pi n}{\lambda}\right)^2}}.
\end{align}

The preceding calculation illustrates how a paraxial Gaussian beam can be characterized through a single measurement of the input field's intensities and relative phases at five pixel positions.
We have derived an analytical description that allows us to determine the displacement of the beam relative to the chip, the beam's inclination, the position of the focal point, and the size of the focal spot.
\newpage
\section{Chip design and bandwidth considerations}
Here, we discuss the design of the main components of the integrated circuit, namely the waveguide cross\-/section, the Y\-/branch, the surface grating couplers, and the passive phase shifters.
\subsection{Waveguide cross-section}
\begin{figure}[H]
     \centering
     \begin{subfigure}[b]{0.48\textwidth}
         \centering
         \caption{}
         \includegraphics[width=\textwidth]{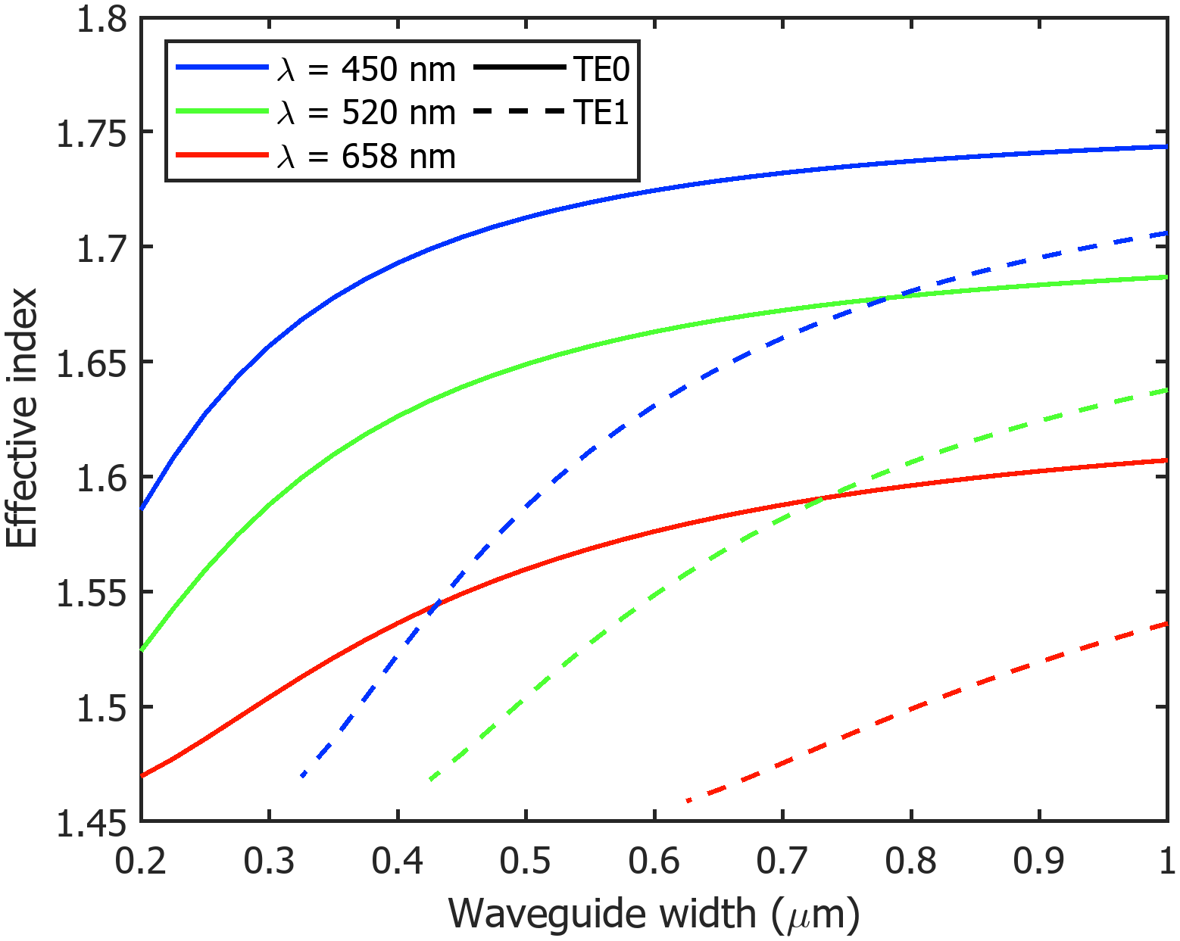}       
     \end{subfigure}
     \hfill
     \begin{subfigure}[b]{0.49\textwidth}
         \centering
         \caption{}
         \includegraphics[width=\textwidth]{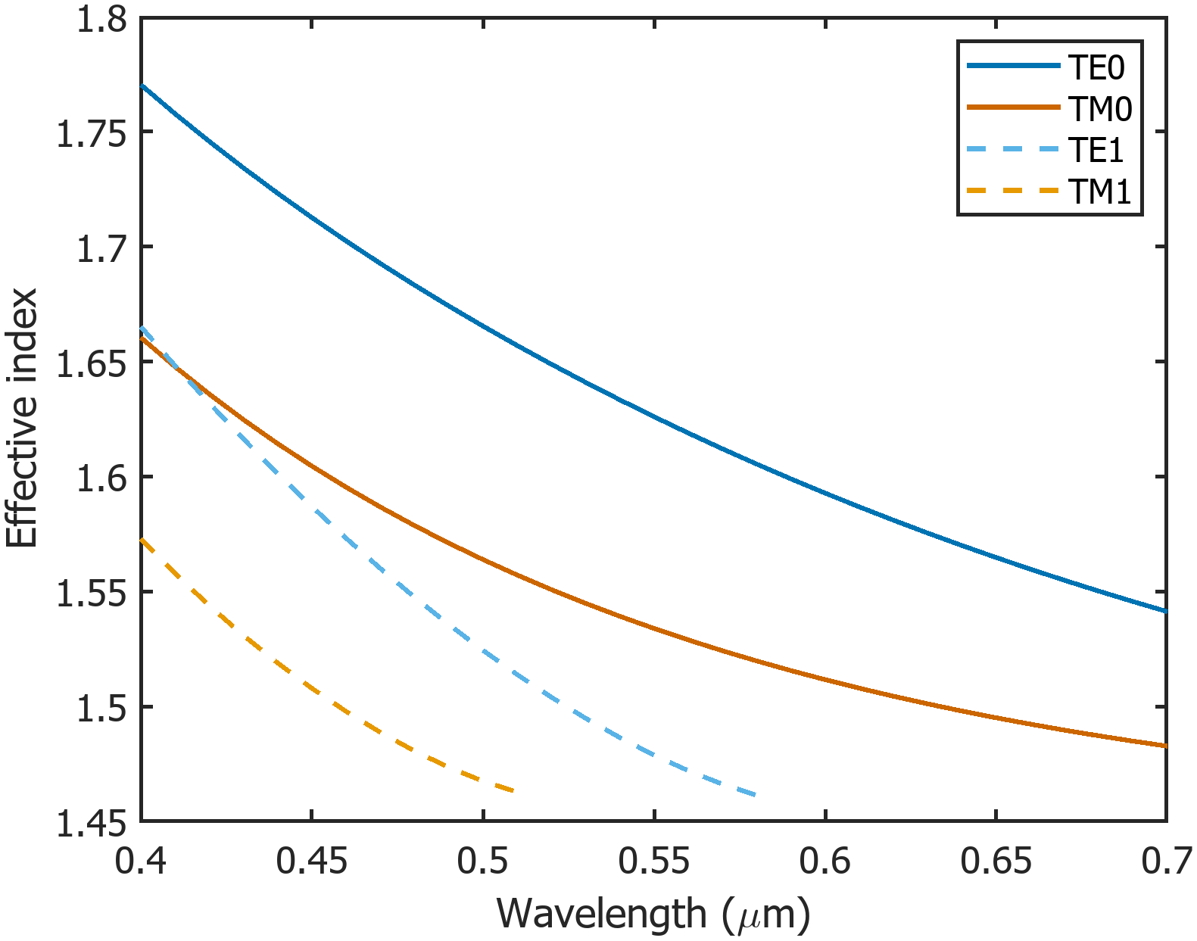}
     \end{subfigure}
        \caption{(a) Effective index of fundamental and first order TE modes at three different wavelengths within the visible spectrum as a function of the waveguide width in our SiN platform. (b) Effective index of fundamental and first order TE and TM modes as a function of wavelength for a waveguide width of 500 nm.}
        \label{fig: Bandwidth}
\end{figure}

The integrated circuit was designed and fabricated in a 100\-/nm\-/thick silicon nitride (SiN) platform with a silicon dioxide bottom cladding and a silica-like top cladding (see Sec. \ref{sec: fabrication} for fabrication details).
The effective index of fundamental and higher\-/order modes calculated through finite difference eigenmode (FDE) simulations at different wavelengths across the visible spectral range shows that the waveguide width must be kept below 600 nm to ensure single\-/mode propagation at the design wavelength of 658 nm (see Fig. \ref{fig: Bandwidth}). To enable circuit operation at shorter wavelengths while maintaining good modal confinement at the design wavelength, a waveguide width of 500 nm was used for the integrated systems, allowing for single-mode operation down to a wavelength of 580 nm. 

\subsection{Grating Couplers}

The single\-/etch self\-/focusing surface grating couplers used in the integrated device were designed to optimize coupling of free\-/space light to the fundamental TE waveguide mode. By simulating the transmission spectrum for different grating periods and fill factors, a period of 525 nm with a fill factor 56\% was chosen to maximize coupling at a wavelength of 658 nm, providing a theoretical 25.6\% input coupling efficiency when considering a 4\-/micron\-/waist Gaussian beam impinging on the surface grating with an incident angle of 12 degrees to the chip surface normal. Angled incidence coupling was chosen to provide higher coupling efficiencies without affecting the chip functionality.  Furthermore, it is possible to couple light at different wavelengths by changing the angle of the incident light to the chip surface normal, as shown in Fig. \ref{fig: Grating Coupler}.

\begin{figure}[H]
     \centering
     \begin{subfigure}[b]{0.49\textwidth}
         \centering
         \caption{}
         \includegraphics[width=\textwidth]{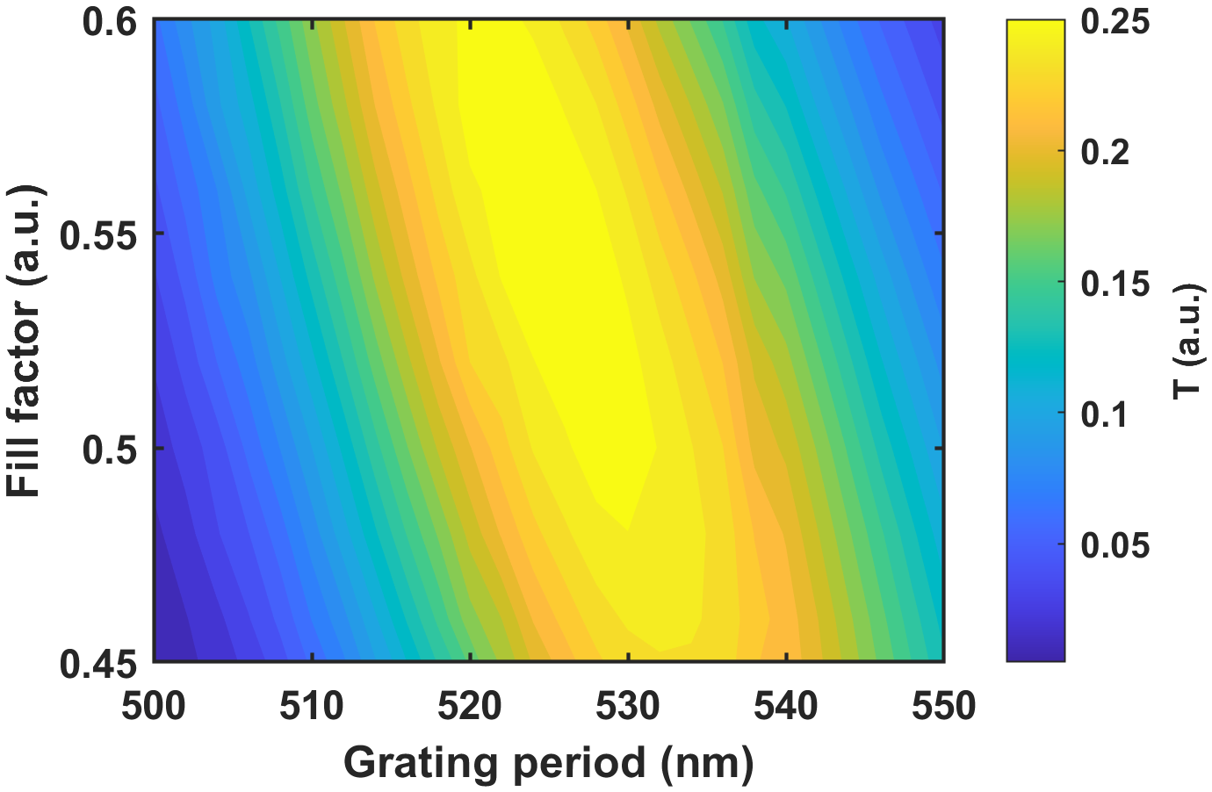}       
     \end{subfigure}
     \hfill
     \begin{subfigure}[b]{0.49\textwidth}
         \centering
         \caption{}
         \includegraphics[width=\textwidth]{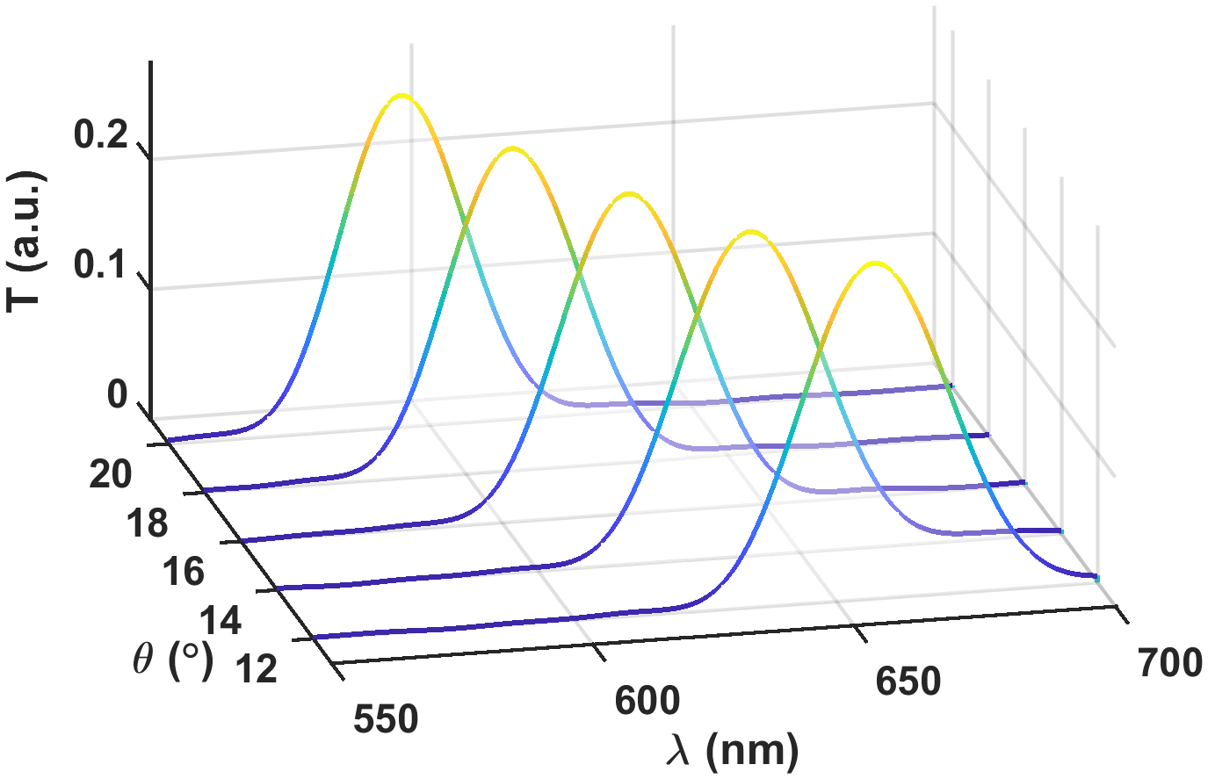}
     \end{subfigure}
        \caption{(a) Grating coupler input coupling efficiency at a wavelength of 658 nm as a function of the grating period and fill factor for a fixed input angle of 12 degrees to the surface normal. (b) Wavelength shift of the grating transmission peak as a function of the input angle.}
        \label{fig:  Grating Coupler}
\end{figure}

\subsection{Y-branch}
A Y\-/branch consists of a waveguide symmetrical splitting in two, resulting in equal power splitting across the two output waveguides. The Y-branches used in this work feature an initial section tapering from 500 nm to 1000 nm width to accommodate the two output waveguides, followed by a short straight section maintaining a width of 1000 nm, from which the two output waveguides branch out. 

The output branch length and final separation (30 µm and 4 µm, respectively) were optimized to minimize insertion loss via a small branching angle, while ensuring a compact footprint. Similarly, the lengths of the taper and subsequent straight sections were optimized to minimize insertion loss across the entire visible spectral range, as shown in Fig. \ref{fig: Y-branch}. The device operates in two different regimes: at longer wavelengths, it behaves as a pure Y-branch, showing relatively stable insertion loss as the length of the taper is changed; at shorter wavelengths, an oscillation of the insertion loss with varying taper length can be observed instead, similarly to what is observed in multimode interference (MMI) devices \cite{Yi:23}. A taper length of 2.5 µm and a straight section length of 1 µm were chosen for broadband minimal insertion loss (< 0.1 dB) across the visible spectral region.
\begin{figure}[H]
     \centering
     \begin{subfigure}[b]{0.59\textwidth}
         \centering
         \caption{}
         \includegraphics[width=\textwidth]{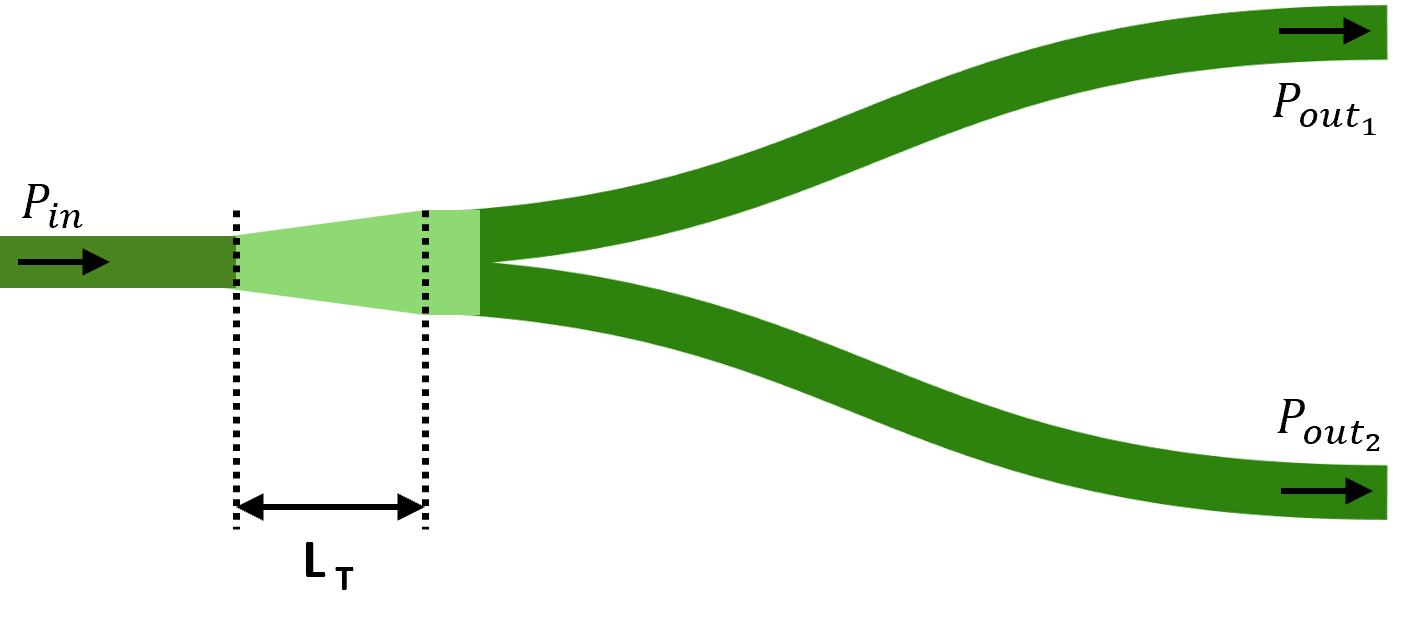}       
     \end{subfigure}
     \hfill
     \begin{subfigure}[b]{0.355\textwidth}
         \centering
         \caption{}
         \includegraphics[width=\textwidth]{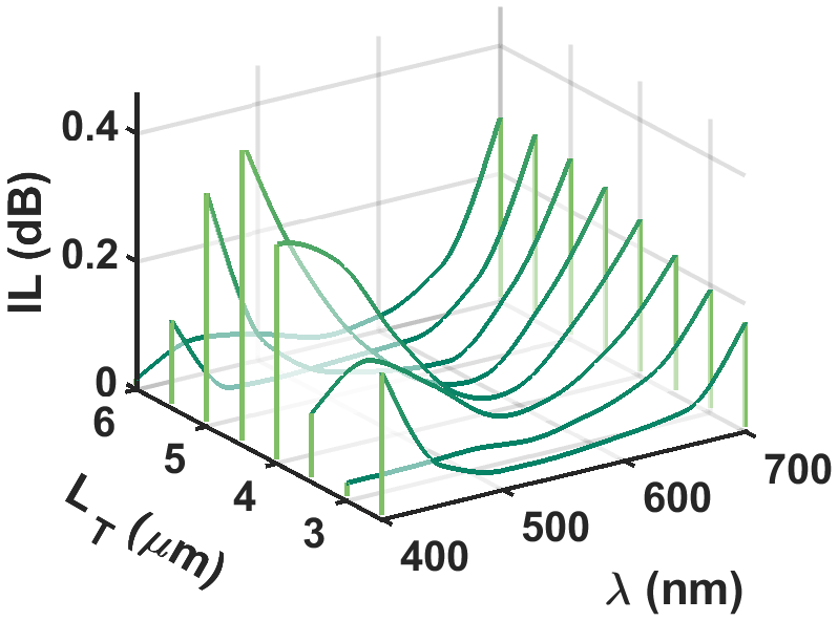}
     \end{subfigure}
        \caption{(a) Schematic of Y\-/branch with taper section highlighted in light green. (b) Simulated insertion loss (IL) across the visible spectral range for different lengths of the taper section $\mathrm{L_T}$.}
        \label{fig:  Y-branch}
\end{figure}
\subsection{Phase shifter}

The passive phase shifters were realized by introducing a difference in the optical path of two waveguide branches of different widths. The resulting difference in effective index introduces a phase shift between the signals at the output of the two branches that is proportional to the difference of the phase constant and the propagation length \cite{Gonzalez-Andrade:20}.

The widths of the two branches were chosen to be 1700 nm and 1900 nm, resulting in 90 degree phase shift for a 93\-/µm\-/long section, and a 135\-/degree phase shift for a 140\-/µm\-/long section at a wavelength of 658 nm. The transition between the 500\-/nm wide single\-/mode waveguide and the wider section was realized through 50\-/µm\-/long tapers to ensure adiabatic mode propagation. However, the phase shifters exhibited random deviations from the target phase shift, producing repeatable results on a given device but differing between different nominally identical devices.

This behavior is most likely due to random width variability and surface roughness of the waveguide interfaces, which can introduce random phase shifting that builds up to non-negligible values across the relatively long waveguides \cite{8608409,8652563}, thus resulting in unpredictable behavior among different devices. Nonetheless, despite the unpredictability of the phase shifts, the calibration procedure is robust enough to overcome the issue and allows for precise phase-front sensing (see Sec. \ref{sec:calibration}).

\subsection{Bandwidth considerations}

The broadband design of most circuit components, combined with a calibration procedure that compensates for their inherent chromatic behavior, enable the sensor's effective broadband operation. However, the constraint for the measurement principle to work is that the waveguides must strictly support only the fundamental modes; otherwise, the theoretical model of the chip becomes invalid. In Fig. \ref{fig: Bandwidth} it can be observed that the TE1 mode begins to be supported at wavelengths just below 580 nm. Therefore, it can be assumed that issues with the measurement principle may arise below this threshold wavelength.

\section{Alternative chip architecture}

\begin{figure*}[h!]
    \centering
    \includegraphics[width = 1\textwidth]{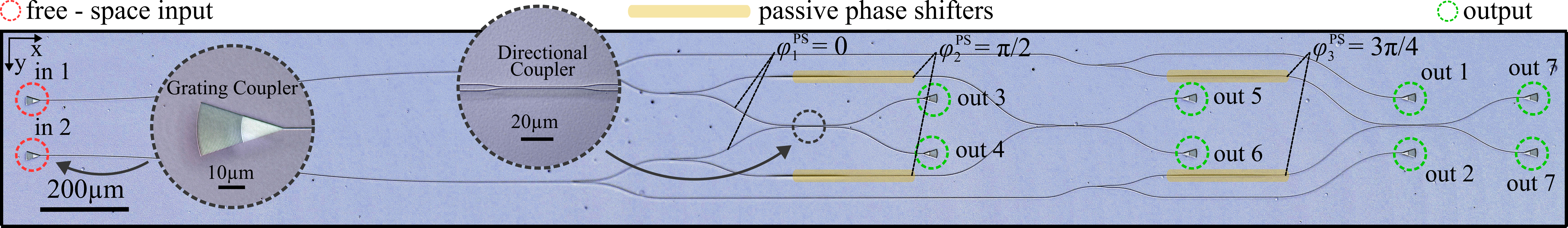}
    \caption[]{Optical microscopy image of a chip featuring an alternative architecture. The on-chip interferometers consist of directional couplers.}
    \label{fig: image-chip-directional-coupler}
\end{figure*}

Here, we present an alternative architecture for the passive on-chip interferometers. 
In this approach, the interferometers are designed using directional couplers instead of Y\-/branch combiners.
Fig. \ref{fig: image-chip-directional-coupler} shows an optical microscope image of a chip featuring directional coupler interferometers.
Throughout this work, chips with this architecture were also calibrated and analyzed.
Both the measurement and calibration methods are applicable to these chips without any conceptual modifications.
In testing, the alternative design exhibited performance comparable to the results discussed in the main text.
However, we believe that Y-branch combiners offer certain advantages for the application presented here, including a smaller footprint and reduced chromatic dispersion.
Therefore, we propose the chip design shown in Fig. \ref{fig: image-chip-directional-coupler} as a substitute, while focusing on the other design in the main text.

\section{Chip Fabrication}
\label{sec: fabrication}

\begin{figure*}[h!]
    \centering
    \includegraphics[width = 1\textwidth]{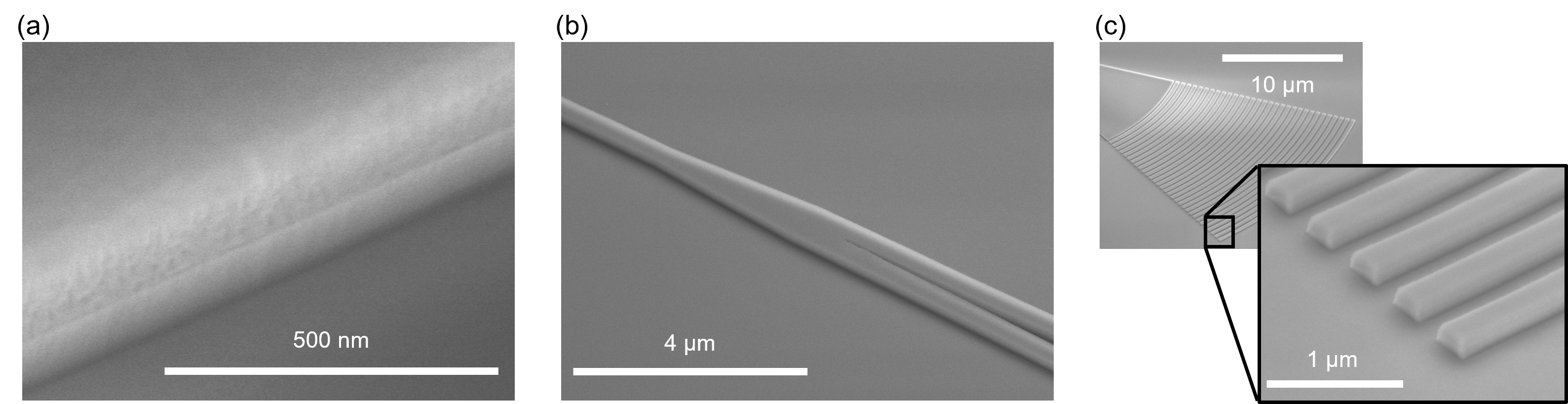}
    \caption[]{(a) SEM image of etched SiN waveguide with residual HSQ mask. (b) SEM image of the fabricated Y-branch. (c) SEM image of the fabricated surface grating coupler.}
    \label{fig: fabrication}
\end{figure*}

The photonic integrated chip was fabricated on a material platform purchased from \textit{LioniX international} and consisting of a 100\-/nm\-/thick SiN film deposited through low\-/pressure chemical vapor deposition on an 8\-/µm\-/thick thermal silicon dioxide layer, mechanically supported by a 500\-/µm\-/thick silicon substrate. A 4\-/inch wafer was diced in 20 mm by 20 mm chips, and the integrated circuit was fabricated through e-beam lithography and plasma etching techniques. 

The integrated circuit was patterned onto the chip by spinning a 270\-/nm\-/thick layer of hydrogen silsesquioxane (HSQ), which was exposed at a dose of 1300 µC/$\mathrm{cm^2}$ and developed in 25\% in tetramethylammonium hydroxide (TMAH). The pattern was then transferred to the SiN film by inductively coupled plasma (ICP) and reactive ion etching (RIE) using a $\mathrm{CHF_3/N_2/O_2}$-based chemistry optimized to ensure good sidewall verticality. Scanning electron microscopy (SEM) images of a waveguide, a Y-branch, and a grating coupler are shown in Fig. \ref{fig: fabrication}. A 500\-/nm\-/thick HSQ layer was finally spun onto the chip and thermally cured to obtain an upper cladding with optical properties close to that of silicon dioxide.

\bibliography{literature}